\documentclass[a4paper,11pt]{article}
\usepackage{pos}

\title{Characterization of the transient response of diamond sensors to collimated, sub-ps, 1 GeV electron bunches}

\author[a]{Silvano~Bassanese}
\author[b]{Luciano~Bosisio}
\author[a,b]{Giuseppe~Cautero}
\author[a,c]{Simone~Di Mitri}
\author[a]{Mario~Ferianis}
\author[c,b]{Alice~Gabrielli}
\author[a,b]{Dario~Giuressi}
\author*[b]{Yifan~Jin}
\author[b]{Livio~Lanceri}
\author[a]{Marco~Marich}
\author[a,b,d]{Ralf~Hendrik~Menk}
\author[c,b]{Lorenzo~Vitale}

\affiliation[a]{Elettra Sincrotrone Trieste SCpA,\\
  AREA Science Park, I-34149 Trieste, Italy}
\affiliation[b]{INFN-Sezione di Trieste,\\
  I-34127 Trieste, Italy}
\affiliation[c]{Dipartimento di Fisica, Universit\`a di Trieste,\\ I-34127 Trieste, Italy}
\affiliation[d]{University of Saskatchewan, \\ S7N Saskatoon, Saskatchewan, Canada.

}

\emailAdd{Yifan.Jin@ts.infn.it}

\abstract{Diamond sensors (DS) are widely used as solid-state particle detectors, beam loss monitors, and dosimeters in high-radiation environments, e.g., particle colliders. We have calibrated our DS with steady $\beta$- and X-radiation, spanning a dose rate in the range 0.1-100 mGy/s. Here, we report the first systematic characterization of transient responses of DS to collimated, sub-picosecond, 1 GeV electron bunches. These bunches, possessing a charge ranging from tens to hundreds of pC and a size from tens of microns to millimeters, are suitably provided by the FERMI electron linac in Trieste, Italy. The high density of charge carriers generated by ionization in the diamond bulk causes a transient modification of electrical properties of DS (e.g., resistance), which in turn affects the signal shape. We have modeled a two-step numerical approach, simulating the effects on the signal of both the evolution of charge carrier density in the diamond bulk and the changes in the circuit parameters. This approach interprets features observed in our experimental results to a great extent.}

\FullConference{%
  41st International Conference on High Energy physics - ICHEP2022\\
  6-13 July, 2022\\
  Bologna, Italy
}

\begin{document}
\maketitle

\section{Introduction}
Diamond sensors (DS) are widely used as solid-state particle detectors, beam loss monitors, and dosimeters in high-radiation environments, e.g., particle colliders~\cite{ref_natural, CANALI1979, ref_atlas, ref_cms, ref_babar}. As a semiconductor material, diamond has high charge carrier mobility and a wide bandgap, which respectively entails fast signal and excellent radiation hardness. Thanks to these features, diamond sensors stand out as a promising candidate for solid-state ionization chambers in radiation dosimetry. 

We have developed and installed~\cite{ref_performance} a diamond-based radiation monitor on the Belle II detector at SuperKEKB electron-positron collider. At the time of writing, SuperKEKB serves as the world record maker (instantaneous luminosity:~$4.7 \times 10^{35}$ cm$^{-2}$s$^{-1}$) at the luminosity frontier. In pursuit of even higher instantaneous luminosity, high radiation bursts, induced by beam loss near the interaction point, which can cause localized damage to essential Belle II sub-detectors and SuperKEKB components, show up as the main bottleneck. As a result, an investigation of the response of diamond sensors to extreme situations is required so as to accurately estimate the beam background severity. In this contribution, we investigate the response of diamond sensors to ultra-fast and intense electron pulses.

\section{Experimental measurements}
\subsection{Diamond Sensor}
The diamond sensors are composed of $(4.5 \times 4.5 \times 0.5)$ mm$^3$ high-purity sCVD diamond crystals from Element Six~\cite{ref_e6} and two $(4.0 \times 4.0)$\,mm$^{2}$ electrodes on opposite faces. The electrodes, which are Ti/Pt/Au layers with $(100$+$120$+$250)$\,nm thickness, are processed by CIVIDEC~\cite{ref_cividec}. We have measured~\cite{ref_cali} the mobility and saturation velocity of charge carriers, and the energy to liberate an electron-hole pair in diamond using $\alpha$ particle source.
In addition, using $\beta$- and X-ray, the stability of DS's response to steady irradiation has been investigated. Also, the calibration factor, from current to dose rate, has been obtained using a silicon diode as a reference, spanning a dose rate in the range 0.1-100 mGy/s.

\subsection{Beam facility}
Free Electron laser Radiation for Multidisciplinary Investigations (FERMI) is the seeded free electron laser (FEL) facility in Trieste, Italy~\cite{ref_FEL}. Its electron linac can provide bunches with a duration of sub-picosecond and a repetition rate from 10 to 50 Hz. The bunch charge can be tuned from tens to hundreds of pC. The energy of the electrons in the bunches ranges from 0.9 GeV to 1.5 GeV with a spread from 0.01\% to 0.3\%. The collimated beam can reach a transverse size down to about 0.1 mm.

\begin{figure}
	\centering
	    \includegraphics[width=15cm]{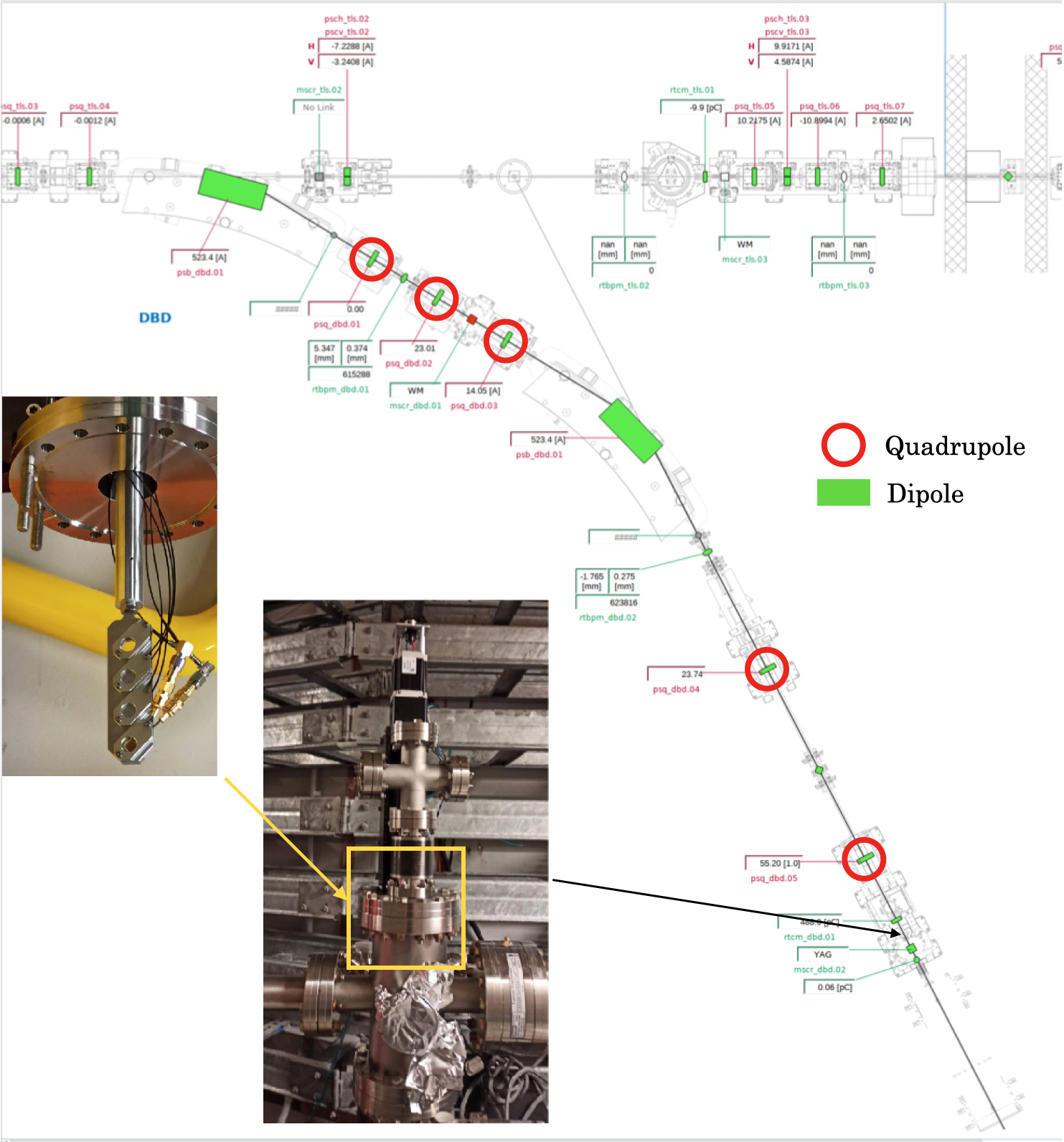}
	\caption{Sketch of the experimental layout. The bending and focusing insertion is composed of two dipole bending magnets (green rectangular) and five quadrupole focusing magnets (red circle). DS is installed in a vacuum chamber located at the end of DBD. Two photographs are of the mechanical support (left) and the vacuum chamber (middle).}
	\label{fig:layout}
\end{figure}

At the end of the linac, before the undulators of FEL, a bending and focusing insertion, composed of two dipole bending magnets and five quadrupole focusing magnets, is installed to deflect the electrons to a diagnostic beam dump (DBD) when activated. A sketch is shown in Figure~\ref{fig:layout}. DS is installed on a mechanical support that can move vertically in a vacuum chamber about one meter downstream of the last quadrupole magnet. A YAG fluorescent screen and a beam current transformer are installed about half a meter downstream of the mechanical support. These two instruments can provide quantitative knowledge of the beam, e.g., its position, transverse size, and charge of the beam bunch, which serves as crucial feedback in beam tuning. 

\subsection{Data taking \& results}
Electrodes of the diamond sensor are connected, via 3-meter coaxial cables, respectively to an HV power supply~\cite{ref_HV} and a LeCroy HDO9000 oscilloscope~\cite{ref_scope} located in a radiation-protected region. During the experiment, the response of the DS is measured under three bias voltages ($V_{bias}$), 50~V, 100~V, and 150~V. The mechanical support is controlled by a stepper motor. By moving the mechanical support, a vertical beam scan is carried out to guarantee that the beam is incident on the center of DS. 

\begin{figure}
	\centering
	    \includegraphics[width=6cm]{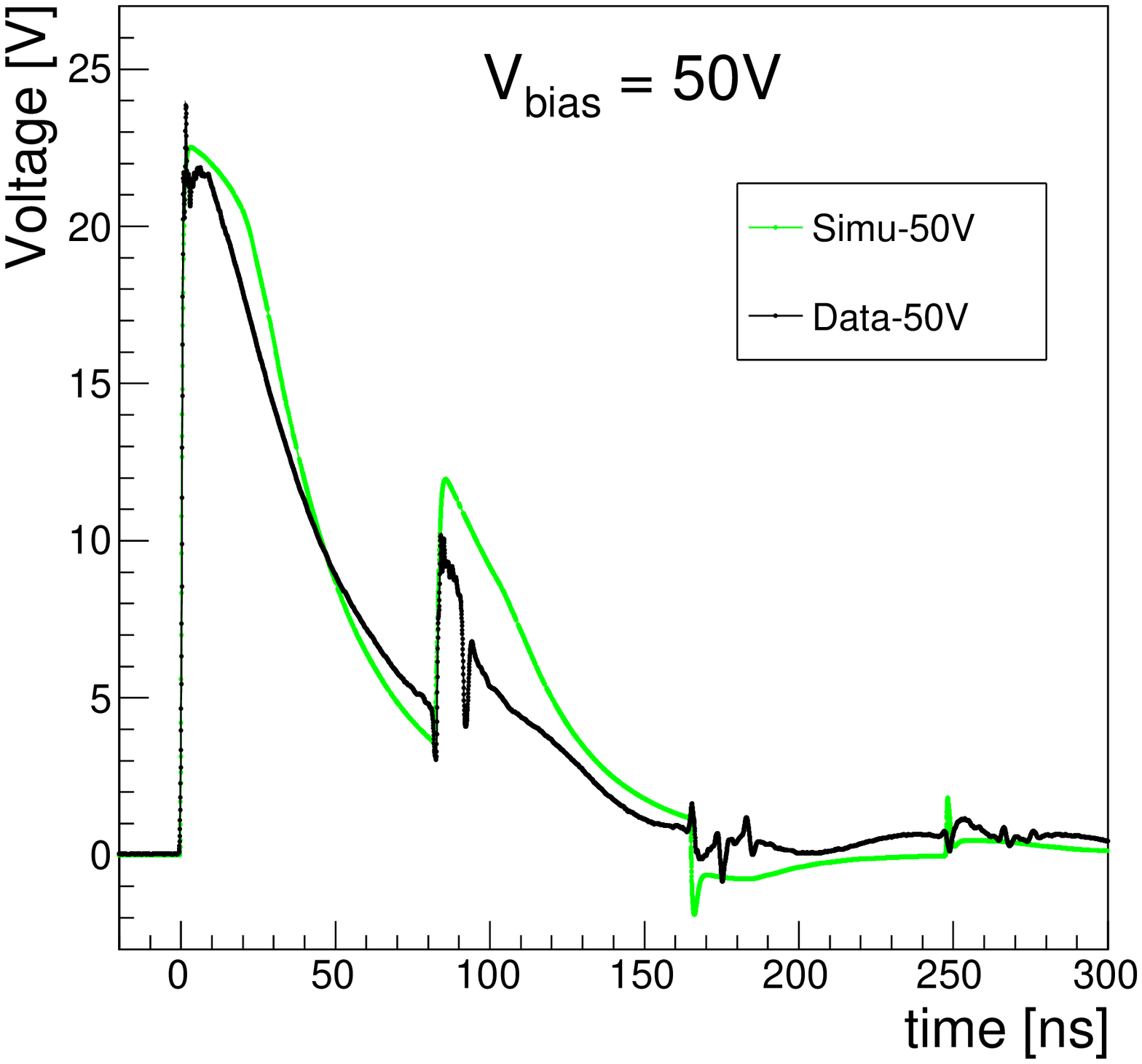}
	    \includegraphics[width=6cm]{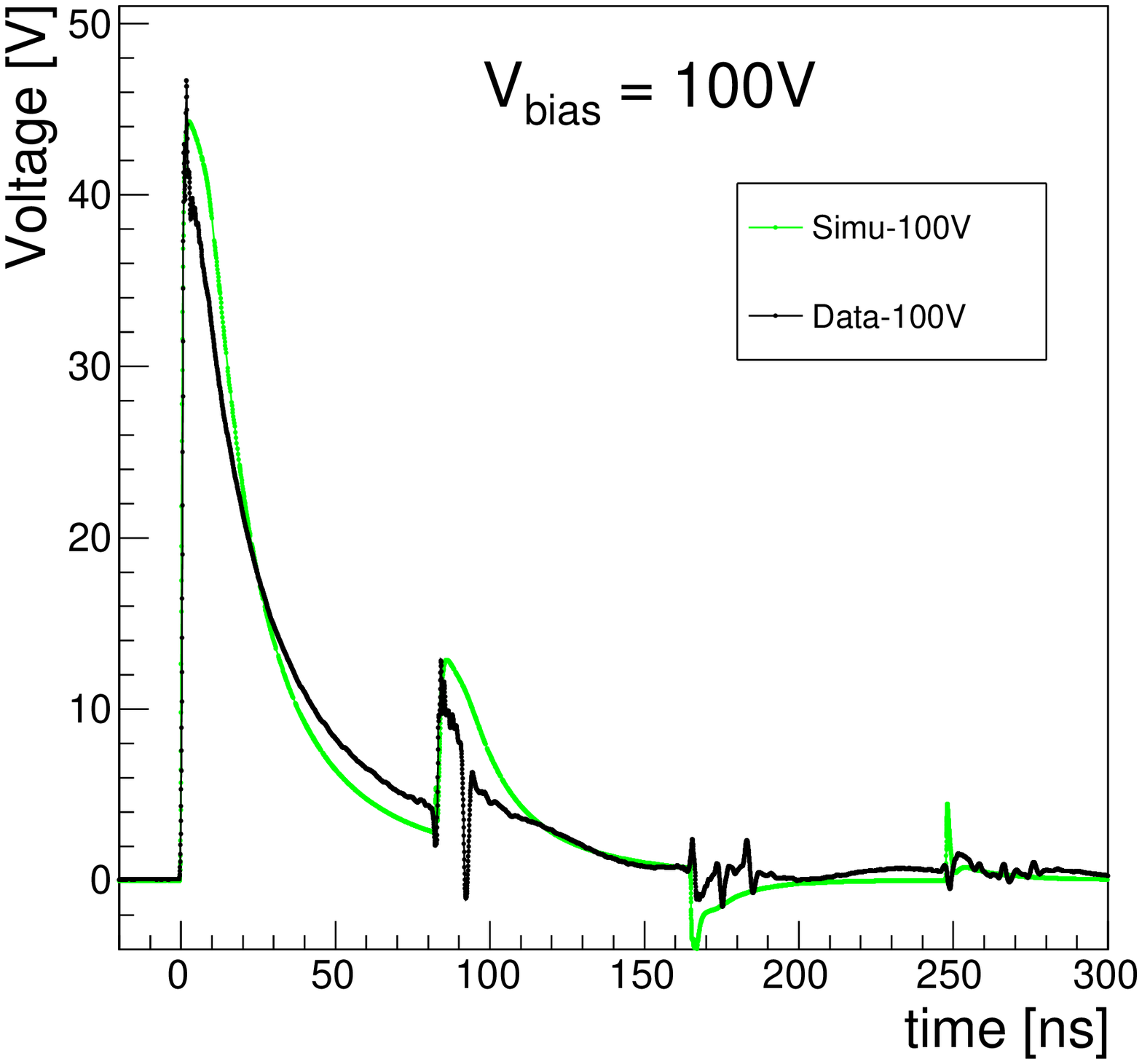}
	    \includegraphics[width=6cm]{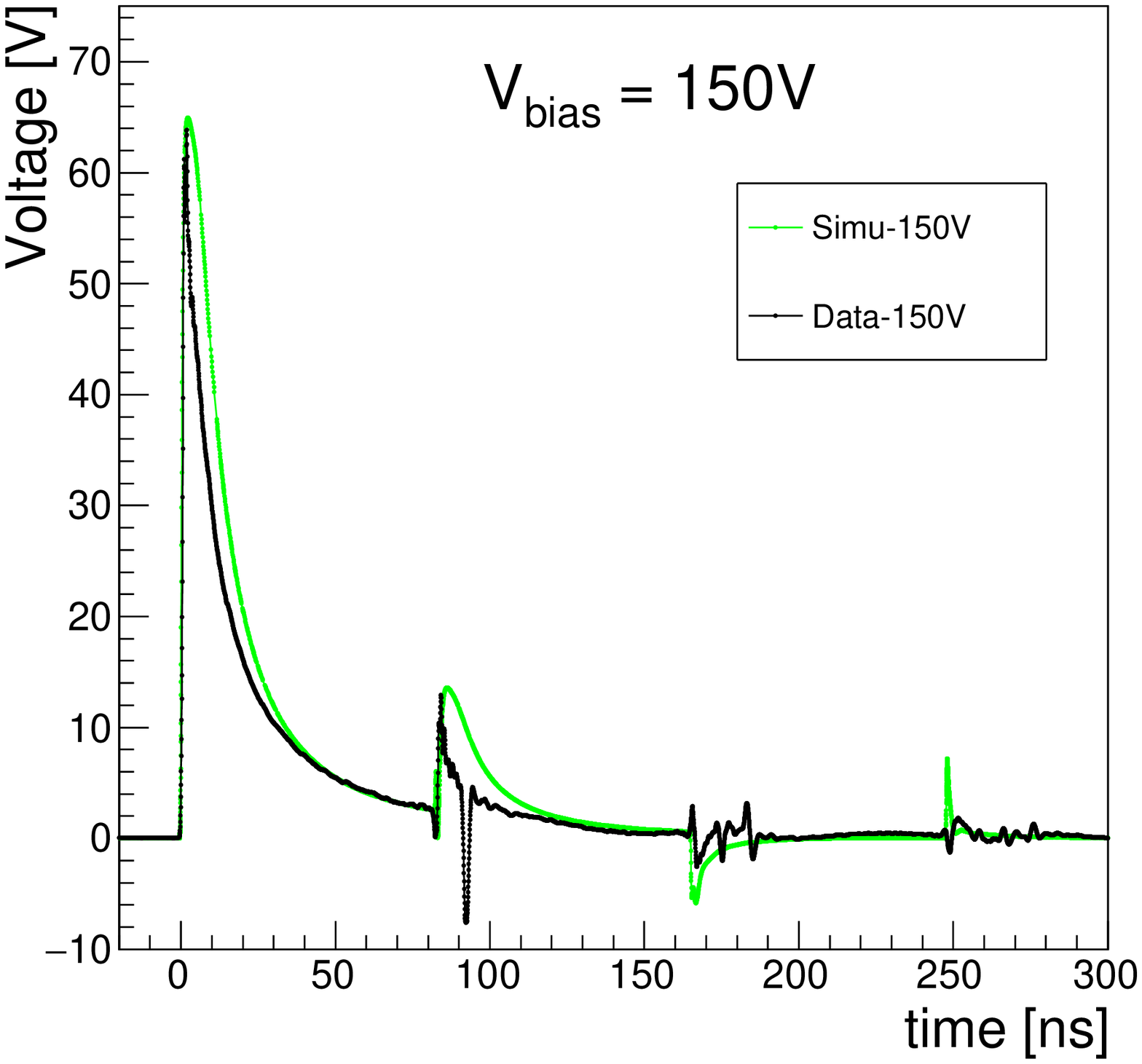}
	\caption{Voltage observed by the oscilloscope as a function of time. Response of the diamond sensor to 35~pC electron bunches with $V_{bias} = $ 50~V, 100~V, and 150~V respectively. Black curves indicate the results of measurements. Green curves indicate the results of the simulation obtained with a lifetime of 50~ns for charge carriers.}
	\label{fig:final}
\end{figure}

After a thorough program of beam tuning, the first data taking has been carried out successfully with a bunch charge 35~pC ($\sim$1\% of the bunch charge at SuperKEKB) and a transverse size around 120 \textmu m. The measured responses of the diamond sensors to the 35~pC electron bunches are shown in Figure~\ref{fig:final} (black curves). Assuming collision energy loss for incoming electrons $\triangle E=0.35$ MeV~\cite{NIST}, the ionization energy of the electron bunch traversing diamond with 0.5~mm thickness can liberate $5.9 \times 10^{12}$ electron-hole pairs, which leads to an expected total signal charge $9.4 \times 10^{-7}$ C. A measure of experimentally collected charge is obtained by integrating the voltage in Figure~\ref{fig:final} and dividing it by 50~$\Omega$ impedance, which gives $2.7 \times 10^{-8}$ C ($V_{bias}=50$~V), $3.0 \times 10^{-8}$ C ($V_{bias}=100$~V), and $2.4 \times 10^{-7}$ C ($V_{bias}=150$~V). The experimentally collected charges account for only $\sim$3\% of the expected value, where a salient non-linearity is manifest.

\section{Interpretation from simulation}
A two-step simulation (TCAD-Sentaurus + LTspice) has been established to interpret the experimental data. Using TCAD Sentaurus~\cite{ref_TCAD}, a series of processes have been simulated, including a beam interacting with the diamond bulk, the liberation of electron-hole pairs, the drift of charge carriers, and the evolution of the induced voltage drop on the electrodes. After that, the simulated voltage drop on the electrode is input to LTspice~\cite{ref_ltspice}, in which an equivalent circuit of the system is drawn. The DS is implemented by a combination of a voltage source in series with a resistor and in parallel with a capacitor. The voltage source used the simulated evolution of voltage from TCAD and the resistance is modeled by the voltage across the diamond electrodes divided by the current in TCAD. Coaxial cables, power supply, and oscilloscope are all modeled to take into account the transmission effects on the electrical signal such as reflection, attenuation, and distortion. 

To precisely describe the physical property of our DS, values of parameters, such as mobility and saturation velocity of charge carriers and the energy to create an electron-hole pair in the diamond that are measured in Ref.~\cite{ref_cali}, are adopted in the simulation. The results of the simulation are shown in Figure~\ref{fig:final} using green curves. A Fair agreement is observed between the results of the simulation and that of experimental measurements.

The underlying mechanism of the non-linear charge collection is revealed by the simulation as follows. The charge carriers generated by ionization in the diamond bulk own an extremely high density ($\sim$10$^{17}$ cm$^{-3}$) that can form an internal electric field sufficient to cancel off the external electric field. This screening effect (a.k.a. plasma effect) delays the collection of charge carriers at the electrodes. In addition, these high-density charge carriers cause a transient modification of the electrical properties of DS, e.g., its impedance. The impedance of each component in the circuit (DS, power supply, and oscilloscope) play an important role in determining the amplitude of the signal. The secondary peaks in the signal are electrical reflections due to impedance mismatch. The impedance of the diamond detector and that of the HV power supply mismatch with 50~$\Omega$ cables. Thus the signal is bouncing back and forth between the diamond detector and the HV power supply.

~~

\section{Conclusion}

We report the first systematic characterization of transient responses of DS to collimated, sub-picosecond, 1 GeV electron bunches. Salient non-linearity of charge collection is observed in DS's response. The underlying mechanism, originating from the screening effect, has been revealed by the simulation. The experimental data collected with 35~pC bunches and the numerical simulation can extrapolate DS's performance to similar radiation conditions. 



\end{document}